\documentclass[superscriptaddress,eqsecnum,floatfix,aps,physrev,repring,twocolumn]{revtex4}
\usepackage{graphicx}
\usepackage{amsmath}
\usepackage{amssymb}
\usepackage{bm}

\begin{document}

\title{Effect of charge renormalization on electric and thermo-electric transport\\
along the vortex lattice of a Weyl superconductor}
\author{G. Lemut}
\affiliation{Instituut-Lorentz, Universiteit Leiden, P.O. Box 9506, 2300 RA Leiden, The Netherlands}
\author{M. J. Pacholski}
\affiliation{Instituut-Lorentz, Universiteit Leiden, P.O. Box 9506, 2300 RA Leiden, The Netherlands}
\author{\.{I}. Adagideli}
\affiliation{Faculty of Engineering and Natural Sciences, Sabanci University, Orhanli-Tuzla, 34956 Istanbul, Turkey}
\author{C. W. J. Beenakker}
\affiliation{Instituut-Lorentz, Universiteit Leiden, P.O. Box 9506, 2300 RA Leiden, The Netherlands}
\date{April 2019}
\begin{abstract}
Building on the discovery that a Weyl superconductor in a magnetic field supports chiral Landau level motion along the vortex lines, we investigate its transport properties out of equilibrium. We show that the vortex lattice carries an electric current $I=\tfrac{1}{2}(Q_{\rm eff}^2/h)(\Phi/\Phi_0) V$ between two normal metal contacts at voltage difference $V$, with $\Phi$ the magnetic flux through the system, $\Phi_0$ the superconducting flux quantum, and $Q_{\rm eff}<e$ the renormalized charge of the Weyl fermions in the superconducting Landau level. Because the charge renormalization is energy dependent, a nonzero thermo-electric coefficient appears even in the absence of energy-dependent scattering processes.
\end{abstract}
\maketitle

\section{Introduction}
\label{intro}

Weyl superconductors are nodal superconductors with topological protection \cite{Men12,Sch15}: They have nodal points of vanishing excitation gap, just like \textit{d}-wave superconductors \cite{Har95}, but in contrast to those the gapless states are not restricted to high-symmetry points in the Brillouin zone and can appear for conventional \textit{s}-wave pairing. The nodal points (Weyl points) at $\pm K$ in a Weyl superconductor are protected by the conservation of a topological invariant: the Berry flux of $\pm 2\pi$ at Weyl points of opposite chirality \cite{Nie83,Tur13}.

The distinction between symmetry and topology has a major consequence for the stability of Landau levels in a magnetic field. While in a \textit{d}-wave superconductor the strong scattering of nodal fermions by vortices in the order parameter prevents the formation of Landau levels \cite{Fra00}, in a Weyl superconductor an index theorem for chiral fermions protects the zeroth Landau level from broadening \cite{Pac18}. The appearance of chiral Landau levels in a superconducting vortex lattice produces a quantized thermal conductance parallel to the magnetic field, in units of $1/2$ times the thermal quantum per $h/2e$ vortex \cite{Pac18}. The factor of $1/2$ reminds us that Bogoliubov quasiparticles are Majorana fermions, ``half a Dirac fermion'' \cite{Bee14,Fra15}. 

\begin{figure}[tb]
\centerline{\includegraphics[width=1\linewidth]{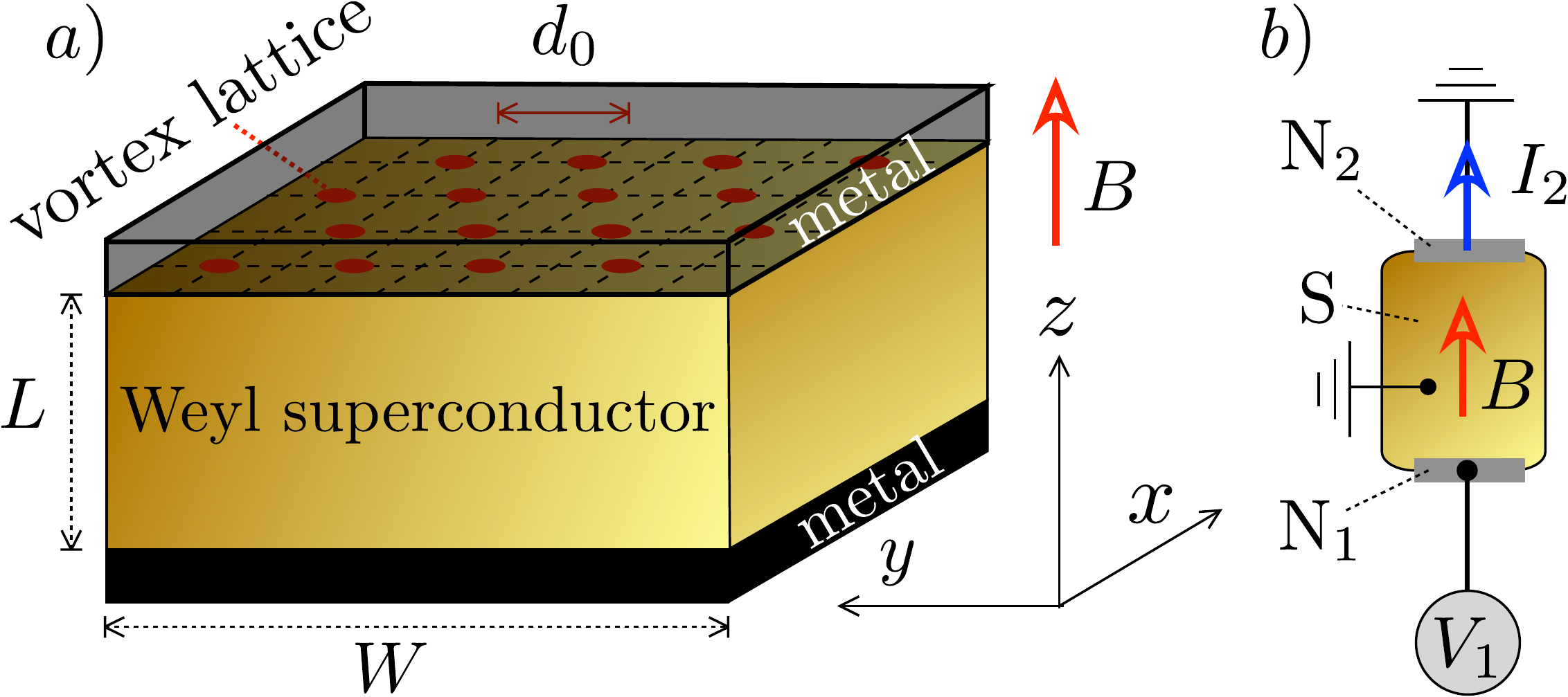}}
\caption{a)  Vortex lattice in a Weyl superconductor sandwiched between metal electrodes; b) Circuit to measure the electrical transport along the vortex lines. The nonlocal conductance $G_{12}=dI_2/dV_1$ gives the current carried through the vortex lattice by nonequilibrium Weyl fermions in a chiral Landau level.
}
\label{fig_layout}
\end{figure}

In this paper we turn from thermal transport to electrical transport, by studying the geometry of Fig.\ \ref{fig_layout} and addressing the question ``What is the charge transported along the vortices in a chiral Landau level?'' It is known \cite{Bai17} that the charge of Weyl fermions in a superconductor (pair potential $\Delta_0$) is reduced by a factor $\kappa=K(\Delta_0)/K(0)$. We find a direct manifestation of this charge renormalization in the electrical conductance, which is quantized at $\tfrac{1}{2}(e\kappa)^2/h$ per vortex. Because the charge renormalization is energy dependent, a coupling between thermal and electrical transport appears even without any energy-dependent scattering mechanism --- resulting in a nonzero thermo-electric effect in a chiral Landau level.

In the next section \ref{sec_LL} we summarize the effective low-energy theory of the superconducting vortex lattice \cite{Pac18}, on which we base our scattering theory in Sec.\ \ref{sec_transmission}, followed by a calculation of electrical and thermo-electric transport properties in Sec.\ \ref{sec_transport}. These analytical results are compared with numerical simulations of a tight-binding model in Sec.\ \ref{sec_numerical}. We conclude in Sec.\ \ref{sec_conclude}.

\section{Landau level Hamiltonian in the vortex lattice}
\label{sec_LL}

\begin{figure*}[tb]
\includegraphics[width=0.45\linewidth]{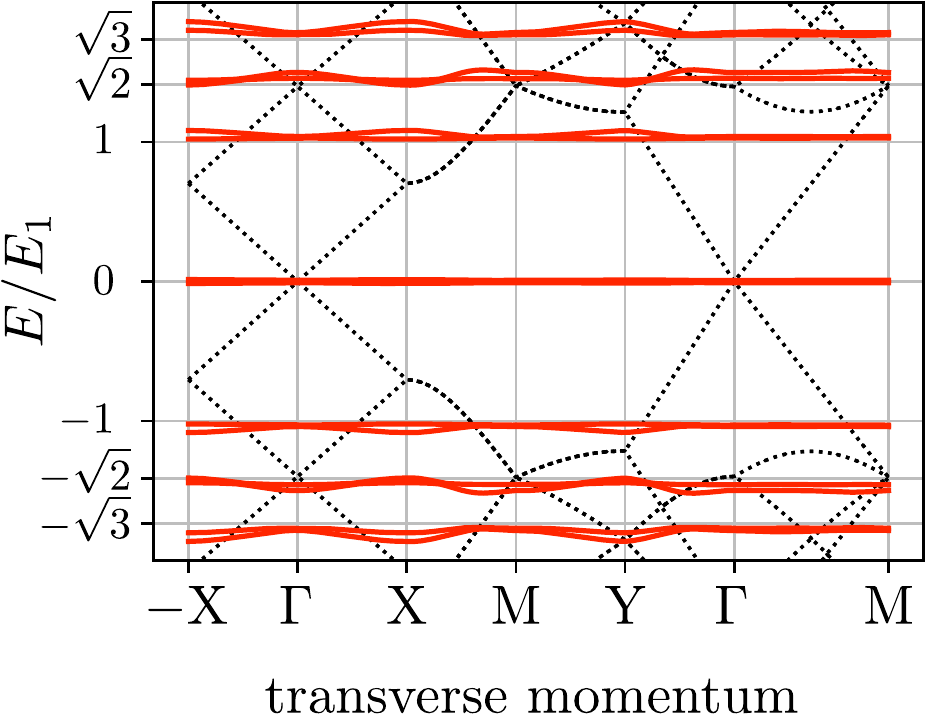}\hfill
\includegraphics[width=0.45\linewidth]{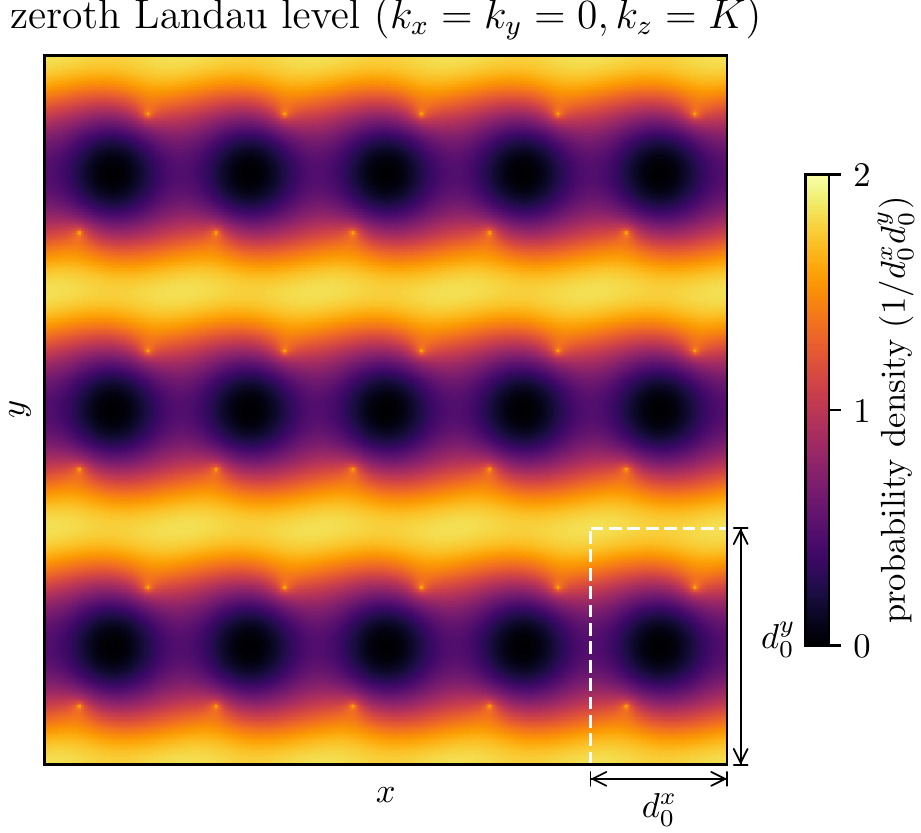} 
\caption{\textit{Left panel:} The red solid curves show the dispersion of Landau levels in the $k_x$--$k_y$ plane perpendicular to the magnetic field (energy $E$ normalized by the energy $E_1$ of the first Landau level). The black dotted curves show the dispersion in zero magnetic field, with a Weyl cone at the $\Gamma$ point of the magnetic Brillouin zone. \textit{Right panel:} Particle density profile in the zeroth Landau level, in the $x$--$y$ plane perpendicular to the magnetic field, for a wave vector at the Weyl point ($\bm{k}=K\hat{z}$). The magnetic unit cell is indicated by a white dashed rectangle. Both panels are calculated numerically for a Weyl superconductor with a triangular vortex lattice. The vortex cores are located at the bright points in the density profile. Similar plots for a square vortex lattice are in Ref.\ \onlinecite{Pac18}.}
\label{fig_triangle}
\end{figure*}

We summarize the findings of Ref.\ \onlinecite{Pac18} for the Landau level Hamiltonian of Weyl fermions in a superconducting vortex lattice, which we will need to calculate the transport properties.

\subsection{Dispersion relation}
\label{sec_dispersion}

\begin{figure}[tb]
\centerline{\includegraphics[width=0.8\linewidth]{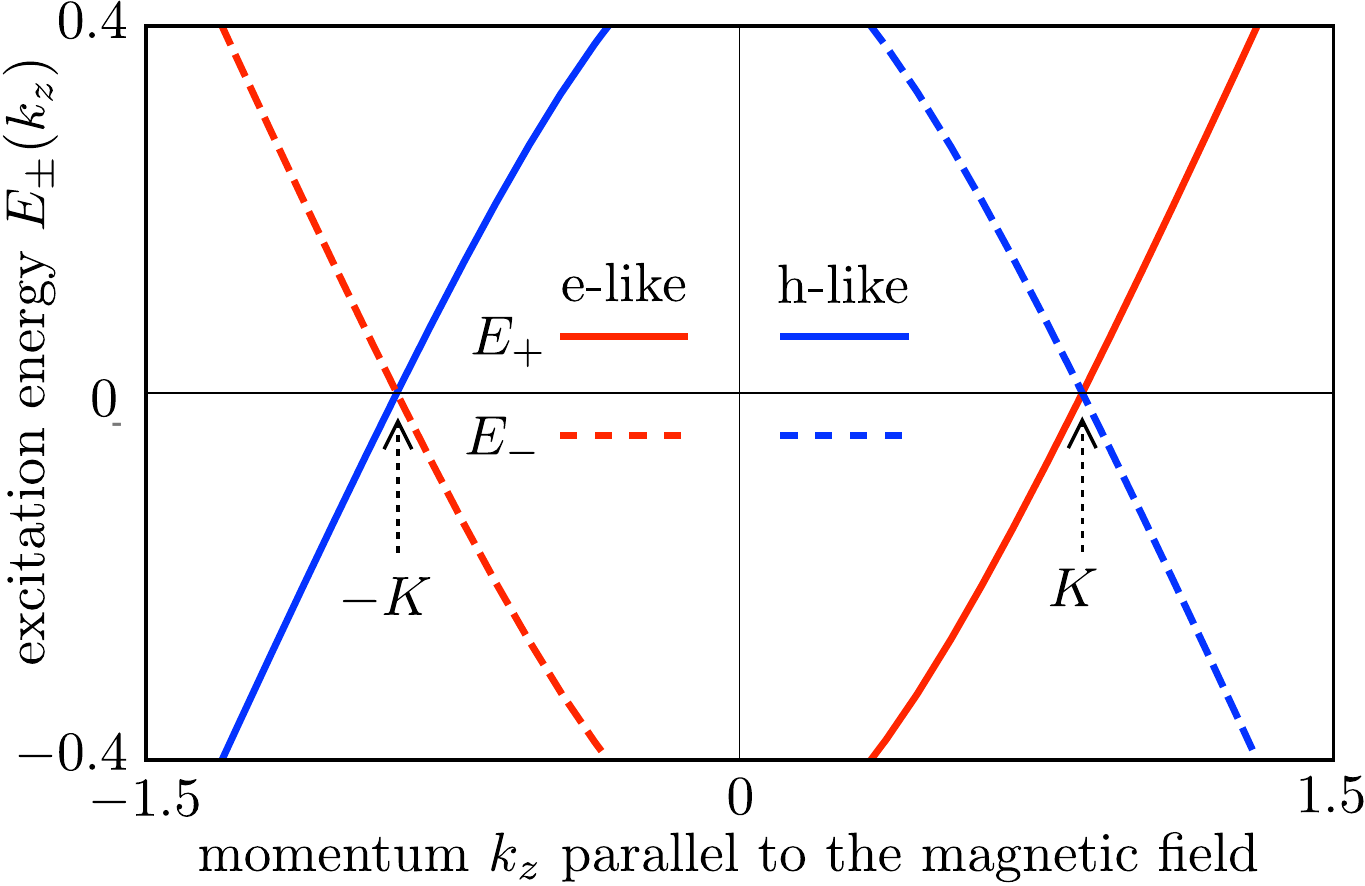}}
\caption{Dispersion relation of the zeroth Landau level in a superconducting vortex lattice, plotted from Eq.\ \eqref{Edispersion} for $\mu=0$, $\Delta_0=0.5$, $\beta=1$. Only the dependence on the momentum $k_z$ along the magnetic field $B$ is shown, the dispersion is flat in the $x$--$y$ plane (see Fig.\ \ref{fig_triangle}). The four branches are distinguished by the sign of the chirality (solid or dashed) and by the sign of the electric charge (red or blue). The zero-field Weyl points at $k_z=\pm K$ are indicated by arrows. Each branch has a degeneracy $N_{\rm Landau}=e\Phi/h$ set by the enclosed flux $\Phi=BW^2$.
}
\label{fig_LL}
\end{figure}

A Landau level is a dispersionless flat band in the plane perpendicular to the magnetic field. The lowest (zeroth) Landau level is protected by chiral symmetry from scattering by the vortices, see Fig.\ \ref{fig_triangle}. This is the Landau level on which we focus our analysis. It is a celebrated result of Nielsen and Ninomiya \cite{Nie83} that Weyl fermions in the zeroth Landau level have a definite chirality $\chi=\pm 1$, defined as the sign of the velocity $v_z=\partial E/\partial k_z$, parallel or antiparallel to $B$. To account for the electron-hole degree of freedom the number of bands is doubled for each chirality, so that we have four bands in total. Electron-like and hole-like bands are related related by the charge-conjugation symmetry relation $E_\chi(k_z)=-E_\chi(-k_z)$.

The effect of a superconducting vortex lattice on this four-band dispersion is given by \cite{Pac18}
\begin{equation}
\begin{split}
&E_\chi(k_z)=-(\text{sgn}\,k_z)\chi M(k_z)-\chi\mu\kappa(k_z),\\
&M(k_z)=\beta-\sqrt{\Delta_0^2+ k_z^2},\;\;\kappa(k_z)=\frac{d}{dk_z}M(k_z),
\end{split}\label{Edispersion}
\end{equation}
plotted in Fig.\ \ref{fig_LL}. (We have set $\hbar$ and the Fermi velocity $v_{\rm F}$ equal to unity, so $\kappa$ is dimensionless.) The magnitude of the superconducting pair potential outside of the vortex cores is denoted by $\Delta_0$ and $\beta$ is an internal magnetization along the $z$-direction that breaks time-reversal symmetry even in the absence of any external magnetic field. In Eq.\ \eqref{Edispersion} we have assumed that $\beta$ is parallel to $B$, but we will later relax this assumption (see Sec.\ \ref{sec_isotropy}).

Provided that $\Delta_0<\beta$ there is a pair of Landau levels for each chirality, located in the magnetic Brillouin zone near the Weyl points at $k_z= K$ and $k_z=-K$, with \cite{Men12}
\begin{equation}
K(\Delta_0)=\sqrt{\beta^2-\Delta_0^2}.\label{Kzdef}
\end{equation}
The charge expectation value 
\begin{equation}
Q_\chi=-e\frac{\partial E_\chi}{\partial\mu}=e\chi\kappa(k_z)=-\frac{e\chi k_z}{\sqrt{\Delta_0^2+k_z^2}}\label{Qchidef}
\end{equation}
for a given chirality has the opposite sign at the two Weyl points. (We say that the chiral Landau levels near $k_z=\pm K$ are charge-conjugate.) When $k_z=\pm K$ is at the Weyl point, the charge renormalization factor equals $\mp\kappa_0$, with
\begin{equation}
\kappa_0=K(\Delta_0)/K(0)=\sqrt{1-\Delta_0^2/\beta^2},\label{kappa0def}
\end{equation}
while $\kappa(k_z)$ varies linearly with energy away from the Weyl point \cite{Bai17}.

\subsection{Effective Hamiltonian}
\label{sec_effH}

The dispersion \eqref{Edispersion} follows from the effective low-energy Hamiltonian \cite{Pac18}
\begin{subequations}
\label{Heffdef}
\begin{align}
{\cal H}={}&U\begin{pmatrix}
H_+&0&0&0\\
0&\cdot&\cdot&0\\
0&\cdot&\cdot&0\\
0&0&0&H_-
\end{pmatrix}U^\dagger,\label{Heffdefa}\\
H_\chi={}&(k_x+e{\cal A}_{\chi,x})\sigma_x+(k_y+e{\cal A}_{\chi,y})\sigma_y\nonumber\\
&+M\sigma_z-\chi\mu\kappa\sigma_0,\label{Heffdefb}\\
U={}&\exp(\tfrac{1}{2}i\theta\nu_y\tau_z\sigma_z),\;\;\theta=\arccos\kappa.\label{Heffdefc}
\end{align}
\end{subequations}
The $2\times 2$ Pauli matrices $\nu_\alpha$, $\tau_\alpha$, and $\sigma_\alpha$ (with $\alpha=0$ the corresponding unit matrix) act on, respectively, the electron-hole, orbital, and spin degrees of freedom. The full Hamiltonian ${\cal H}$ is an $8\times 8$ matrix and the $2\times 2$ matrices $H_\pm$ act on the $\sigma$ index in the $\nu=\tau=\pm 1$ sector. 

The central block in Eq.\ \eqref{Heffdefa} indicated by dots refers to higher-lying bands that are approximately decoupled from the low-energy bands. Virtual transitions to these higher bands contribute order $\mu^2$ terms that remove the discontinuity in the derivative $\partial E/\partial k_z$ at $k_z=0$ for $\mu\neq 0$. No such decoupling approximations are made in the numerics of Sec.\ \ref{sec_numerical}.

The gauge field ${\cal A}_\chi(\bm{r})$, dependent on the position $\bm{r}=(x,y)$ in the $x$--$y$ plane, defines the effective magnetic field ${\cal B}_\chi=\nabla\times{\cal A}_\chi$ in the $z$-direction felt by the Weyl fermions in the  lattice of vortices at positions $\bm{R}_n$,
\begin{equation}
{\cal B}_\chi=(1+\chi\kappa)\Phi_0{\sum_n}\delta(\bm{r}-\bm{R}_n)-\chi\kappa B.
\end{equation}
There are $N_{\text{vortex}} = BW^2/\Phi_0$ vortices of flux $\Phi_0=h/2e$ in an area $W^2$ perpendicular to the applied magnetic field $B$, so the spatial average $\int {\cal B}_\chi d\bm{r}=\Phi$ equals the total enclosed flux $\Phi=BW^2$ independent of $\kappa$ or of the lattice of vortices. (In the numerics that follows we will use a square lattice for definiteness.)

\subsection{Zeroth Landau level wave functions}
\label{sec_zerothLL}

As shown in Ref.\ \onlinecite{Pac18}, the Aharonov-Casher index theorem \cite{Aha79,Kat08,Kai09}, together with the requirement that the wave functions are square-integrable at a vortex core, implies that the zeroth Landau level eigenstates $\psi_{\chi}$ of $H_\chi$, which are rank-two spinors, are also eigenstates $|\pm\rangle_\sigma$ of $\sigma_z$,
\begin{equation}
\sigma_z\psi_{\chi}=(\text{sgn}\,Q_\chi)\psi_\chi.\label{eigenvaluesigmaz}
\end{equation}
The eigenvalue is determined by the sign of the effective quasiparticle charge \eqref{Qchidef}. 

It follows that the eigenstates $\Psi_\chi$ of the full Hamiltonian ${\cal H}$, which are rank-eight spinors, have the form
\begin{align}
\Psi_\chi={}&e^{ik_z z}f_\chi(x,y)e^{\tfrac{1}{2}i\theta\nu_y\tau_z\sigma_z}|\text{sgn}\,\chi\rangle_\nu|\text{sgn}\,\chi\rangle_\tau|\text{sgn}\,Q_\chi\rangle_\sigma\nonumber\\
={}&e^{ik_z z}f_\chi(x,y)\biglb[\cos(\theta/2)|\text{sgn}\,\chi\rangle_\nu|\text{sgn}\,\chi\rangle_\tau|\text{sgn}\,Q_\chi\rangle_\sigma\nonumber\\
&-\sin(\theta/2)(\text{sgn}\,Q_\chi)|\!-\!\text{sgn}\,\chi\rangle_\nu|\text{sgn}\,\chi\rangle_\tau|\text{sgn}\,Q_\chi\rangle_\sigma\bigrb].\label{Psichiresult}
\end{align}
The spatial density profile $f_\chi(x,y)$ is peaked at the vortex cores, with a power law decay $|f_\chi|^2\propto \delta r^{-1+|Q_\chi|/e}$ at a distance $\delta r$ from the core \cite{Pac18}. The renormalization of the quasiparticle charge does not affect the degeneracy of the zeroth Landau level: each of the four chiral modes in Fig.\ \ref{fig_LL} has a degeneracy
\begin{equation}
N_{\text{Landau}}=e\Phi/h\label{NLandaudef}
\end{equation}
set by the bare charge $e$.

Although the spatial density profile of these chiral modes is nonuniform, the wave functions extend over the entire $x$--$y$ plane --- they are not exponentially confined to the vortex cores (see Fig.\ \ref{fig_triangle}). This is a qualitative difference between the zeroth Landau level of a Weyl superconductor and zero-modes bound to vortices in topological superconductors \cite{Vol99,Fu08}.

\section{Transmission through the NSN junction}
\label{sec_transmission}

Refering to the geometry of Fig.\ \ref{fig_layout}, we seek the transmission matrix $t_{\text{NSN}}$ for propagating modes of electrons and holes transmitted from the first metal contact $\text{N}_1$ in the region $z<0$, through the Weyl superconductor in the region $0<z<L$, into the second metal contact $\text{N}_2$ in the region $z>L$.

\subsection{Renormalized charge transfer}
\label{sec_effcharge}

We start by examining a single NS interface, to study how a chiral mode in the superconductor injects a renormalized charge into the normal metal. 

On the superconducting side $z<L$ of the NS interface at $z=L$ the incident modes have positive chirality $\chi=+1$. There is a mode $\Psi_{\rm S}$ with perpendicular momentum $k_z$ near $K$ and a mode $\Psi'_{\rm S}$ with $k'_z$ near $-K$. We do not specify the transverse momentum $\bm{k}_\parallel=(k_x,k_y)$, which gives each mode a degeneracy of $N_{\rm Landau}=e\Phi/h$, see Eq.\ \eqref{NLandaudef}.

According to Eq.\ \eqref{Psichiresult}, the spinor structure of the chiral modes is
\begin{equation}
\begin{split}
&\Psi_{\rm S}\propto \cos(\theta/2)|\mbox{++$-$}\rangle_{\nu\tau\sigma}+\sin(\theta/2)|\mbox{$-$+$-$}\rangle_{\nu\tau\sigma},\\
&\Psi'_{\rm S}\propto \cos(\theta'/2)|\mbox{+++}\rangle_{\nu\tau\sigma}-\sin(\theta'/2)|\mbox{$-$++}\rangle_{\nu\tau\sigma}.
\end{split}
\label{PsiKspinor}
\end{equation}
We have abbreviated $|\mbox{$\pm$$\pm$$\pm$}\rangle_{\nu\tau\sigma}=|\pm\rangle_{\nu}|\pm\rangle_{\tau}|\pm\rangle_{\sigma}$ and denote $\theta=\theta(k_z)$, $\theta'=\theta(k'_z)$.

For the normal metal we take the free-electron Hamiltonian
\begin{equation}
H_{\rm N}=\frac{1}{2m}(k^2-k_{\rm F}^2)\nu_z\tau_0\sigma_0,\label{HNdef}
\end{equation}
isotropic in the spin and valley degrees of freedom, in the high Fermi-momentum limit $k_{\rm F}l_m\rightarrow\infty$ when the effect of the magnetic field on the spectrum may be neglected ($l_m=\sqrt{\hbar/eB}$ is the magnetic length).

Because of the large potential step experienced upon traversing the NS interface, the perpendicular momentum $k_z$ is boosted to $+k_{\rm F}$ for the electron component of the state and to $-k_{\rm F}$ for the hole component. A state in N moving away from the NS interface of the form
\begin{subequations}
\label{Psimatch}
\begin{align}
\Psi_{\rm N}\propto{}& e^{ik_{\rm F}(z-L)}\cos(\theta/2)|\mbox{++$-$}\rangle_{\nu\tau\sigma}\nonumber\\
&+e^{-ik_{\rm F}(z-L)}\sin(\theta/2)|\mbox{$-$+$-$}\rangle_{\nu\tau\sigma}\label{Psimatcha}
\end{align}
can be matched to the incident state $\Psi_S$ in S, while the state
\begin{align}
\Psi'_{\rm N}\propto{}& e^{ik_{\rm F}(z-L)}\cos(\theta'/2)|\mbox{+++}\rangle_{\nu\tau\sigma}\nonumber\\
&-e^{-ik_{\rm F}(z-L)}\sin(\theta'/2)|\mbox{$-$++}\rangle_{\nu\tau\sigma}\label{Psimatchb}
\end{align}
\end{subequations}
can be matched to $\Psi'_S$. 

The charge transferred through the interface when $\Psi_{\text S}\mapsto\Psi_{\text N}$ equals the renormalized charge from Eq.\ \eqref{Qchidef},
\begin{equation}
Q_{\rm N}=\langle\Psi_{\rm N}|e\nu_z|\Psi_{\rm N}\rangle=e\cos\theta=e\kappa=\frac{-ek_z}{\sqrt{\Delta_0^2+k_z^2}},\label{QNresult}
\end{equation}
dependent on the perpendicular momentum $k_z$ in S, before the boost to $k_{\rm F}$ in N. When $k_z=K$, this gives 
\begin{equation}
Q_{\rm N}=-e\sqrt{1-\Delta_0^2/\beta^2}=-\kappa_0 e\equiv -Q_{\rm eff}.\label{Qeffdef}
\end{equation}
This is for the transmission $\Psi_{\text S}\mapsto\Psi_{\text N}$ . The other transmission $\Psi'_{\text S}\mapsto\Psi'_{\text N}$ transfers for $k'_z=-K$ a charge $Q'_{\rm N}=+Q_{\rm eff}$.

Similarly, at the opposite NS interface $z=0$ the chiral Landau level modes in S moving away from the interface are matched to incoming states in N of the form
\begin{subequations}
\label{Phimatch}
\begin{align}
\Phi_{\rm N}\propto{}& e^{ik_{\rm F}z}\cos(\theta/2)|\mbox{++$-$}\rangle_{\nu\tau\sigma}\nonumber\\
&+e^{-ik_{\rm F}z}\sin(\theta/2)|\mbox{$-$+$-$}\rangle_{\nu\tau\sigma},\label{Phimatcha}\\
\Phi'_{\rm N}\propto{}& e^{ik_{\rm F}z}\cos(\theta'/2)|\mbox{+++}\rangle_{\nu\tau\sigma}\nonumber\\
&-e^{-ik_{\rm F}z}\sin(\theta'/2)|\mbox{$-$++}\rangle_{\nu\tau\sigma}.\label{Phimatchb}
\end{align}
\end{subequations}

\subsection{Transmission matrix}
\label{sec_transmatrix}

At a given energy $E$ relative to the Fermi level the perpendicular momenta $k_z$ and $k'_z$ of the chiral Landau levels in S moving in the $+z$ direction are determined by the dispersion relation \eqref{Edispersion} with $\chi=+1$. For $\mu=0$ the expressions are simple,
\begin{equation}
k_z=K+(\beta/K)E,\;\;k'_z=-K+(\beta/K)E.\label{kzEdef}
\end{equation}
For any $\mu$, particle-hole symmetry ensures that
\begin{equation}
k_z(E)=-k'_z(-E).\label{kzphsym}
\end{equation}

The Landau level $\Psi_{\rm S}$ propagating from $z=0$ to $z=L$ accumulates a phase $k_z L$, and similarly $\Psi'_{\rm S}$ accumulates a phase $k'_z L$. The full transmission matrix of the NSN junction at energy $E$ can thus be written as
\begin{align}
t_{\text{NSN}}(E)=e^{ik_z L}|\Psi_{\rm N}\rangle\langle\Phi_{\rm N}|+e^{ik'_z L}|\Psi'_{\rm N}\rangle\langle\Phi'_{\rm N}|,\label{tNSNE}
\end{align} 
with $k_z$ and $k'_z$ determined by Eq.\ \eqref{kzEdef}.

We can rewrite Eq.\ \eqref{tNSNE} in the basis of propagating electron modes in the normal metal. In the region $z<0$ one has the basis states
\begin{subequations}
\label{Psiupdowndef}
\begin{align}
&|\Psi_\uparrow\rangle=\begin{pmatrix}
|e\uparrow\rangle\\
|h\uparrow\rangle
\end{pmatrix},\;\;|\Psi_\downarrow\rangle=\begin{pmatrix}
|e\downarrow\rangle\\
|h\downarrow\rangle
\end{pmatrix},\label{Psiupdowndefa}\\
&|e\uparrow\rangle=e^{ik_{\rm F}z}|\mbox{+++}\rangle_{\nu\tau\sigma},\;\;|h\uparrow\rangle=e^{-ik_{\rm F}z}|\mbox{$-$++}\rangle_{\nu\tau\sigma},\nonumber\\
&|e\downarrow\rangle=e^{ik_{\rm F}z}|\mbox{++$-$}\rangle_{\nu\tau\sigma},\;\;|h\downarrow\rangle=e^{-ik_{\rm F}z}|\mbox{$-$+$-$}\rangle_{\nu\tau\sigma},\label{Psiupdowndefb}
\end{align}
\end{subequations}
and similarly for $z>L$ with $k_{\rm F}z$ replaced by $k_{\rm F}(z-L)$.

The transmission matrix is block diagonal in the spin degree of freedom,
\begin{subequations}
\label{tNSNfinal}
\begin{align}
&t_{\text{NSN}}(E)=\begin{pmatrix}
t_\uparrow(E)&0\\
0&t_\downarrow(E)
\end{pmatrix},\label{tNSNfinala}\\
&t_\uparrow=e^{ik'_z L}\begin{pmatrix}
\cos^2(\theta'/2)&-\cos(\theta'/2)\sin(\theta'/2)\\
-\cos(\theta'/2)\sin(\theta'/2)&\sin^2(\theta'/2)
\end{pmatrix}\nonumber,\\
&t_\downarrow=e^{ik_z L}\begin{pmatrix}
\cos^2(\theta/2)&\cos(\theta/2)\sin(\theta/2)\\
\cos(\theta/2)\sin(\theta/2)&\sin^2(\theta/2)
\end{pmatrix}.\label{tNSNfinalb}
\end{align}
\end{subequations}

The $2\times 2$ matrix $t_\uparrow$ acts on the electron-hole spinor $|\Psi_\uparrow\rangle$ and $t_\downarrow$ acts on  $|\Psi_\downarrow\rangle$. We may write this more compactly as
\begin{equation}
\begin{split}
&t_\uparrow=\tfrac{1}{2}e^{ik'_z L}\left(\nu_0+\nu_z e^{-i\theta'\nu_y}\right),\\
&t_\downarrow=\tfrac{1}{2}e^{ik_z L}\left(\nu_0+\nu_z e^{i\theta\nu_y}\right).
\end{split}\label{tcompact}
\end{equation}
These are each rank-one matrices, one eigenvalue equals 0 and the other equals 1 in absolute value. The unit transmission eigenvalue is $N_{\text{Landau}}$-fold degenerate in the transverse momentum $\bm{k}_\parallel$.

At the Fermi level $E=0$ the particle-hole symmetry relation \eqref{kzphsym} implies $k'_z=-k_z$, $\theta'=\pi-\theta$, hence
\begin{equation}
t_{\text{NSN}}(0)=\tfrac{1}{2}e^{-ik_zL\sigma_z}\left(\nu_0-\nu_z \sigma_z e^{i\theta\nu_y}\right).\label{tzeroenergy}
\end{equation}
One verifies that
\begin{equation}
t_{\text{NSN}}(0)=\nu_y\sigma_y t_{\text{NSN}}^\ast(0)\nu_y\sigma_y,\label{tNSNphsymcheck}
\end{equation}
as required by particle-hole symmetry.

\section{Transport properties}
\label{sec_transport}

The transmission matrix allows us to calculate the transport properties of the NSN junction, under the assumption that there is no backscattering of the chiral modes in the Weyl superconductor. To simplify the notation, we write $t$ for the Fermi-level transmission matrix $t_{\rm NSN}(0)$. The submatrices of electron and hole components are denoted by $t_{ee}$, $t_{hh}$, $t_{he}$, and $t_{eh}$. We define the combinations
\begin{subequations}
\label{Tpmdef}
\begin{align}
&{\cal T}_\pm=t^\dagger_{ee}t^{\vphantom{\dagger}}_{ee}\pm t^\dagger_{he}t^{\vphantom{\dagger}}_{he},\\
&{\cal T}_+=\tfrac{1}{2}(\nu_0+\nu_z)t^\dagger t,\;\;{\cal T}_-=\tfrac{1}{2}(\nu_0+\nu_z) t^\dagger\nu_z t.
\end{align}
\end{subequations}

\subsection{Thermal conductance}
\label{sec_Gthermal}

As a check, we first recover the result of Ref.\ \onlinecite{Pac18} for the quantization of the thermal conductance.

The thermal conductance $G_{\rm thermal}=J_{12}/\delta T$ gives the heat current $J_{12}$ transported at temperature $T_0$ from contact $N_1$ to $N_2$ via the superconductor, in response to a small temperature difference $\delta T$ between the contacts. It follows from the total transmitted quasiparticle current,
\begin{equation}
G_{\rm thermal}=\tfrac{1}{2}g_0 N_{\rm Landau}\,{\rm Tr}\,t^\dagger t=g_0 \frac{e\Phi}{h},\label{Gthermalresult}
\end{equation}
with $N_{\rm Landau}=e\Phi/h$ the Landau level degeneracy and $g_0=\tfrac{1}{3}(\pi k_{\rm B})^2(T_0/h)$ the thermal conductance quantum. The factor $1/2$ in the first equation appears because the quasiparticles in the Weyl superconductor are Majorana fermions. It is cancelled by the factor of two from ${\rm Tr}\,tt^\dagger=2$, in view of Eq.\ \eqref{tzeroenergy}.

\subsection{Electrical conductance}
\label{sec_Gelectric}

Referring to the electrical circuit of Fig.\ \ref{fig_layout}b, we consider the electrical conductance $G_{12}=dI_2/dV_1$, given by
\begin{align}
G_{12}={}&\frac{e^2}{h}N_{\rm Landau}\,{\rm Tr}\,{\cal T}_-\nonumber\\
={}&\frac{e^2}{h}N_{\rm Landau}\tfrac{1}{2}\,{\rm Tr}\,(\nu_0+\nu_z)t^\dagger\nu_z t.\label{G12def2}
\end{align}
In the linear response limit $V_1\rightarrow 0$ we substitute $t$ from Eq.\ \eqref{tzeroenergy}, which gives
\begin{equation}
G_{12}(0)=\cos^2\theta\frac{e^2}{h}N_{\rm Landau}=\frac{(e\kappa)^2}{h}\frac{e\Phi}{h}.\label{G12result}
\end{equation}
The conductance quantum $e^2/h$ is renormalized by the effective charge $e\mapsto e\kappa$. At $\mu=0$, when $k_z=K$, the renormalization factor is $\kappa_0^2=(Q_{\rm eff}/e)^2=1-\Delta_0^2/\beta^2$ from Eq.\ \eqref{Qeffdef}. Note that the conductance per $h/2e$ vortex is $\tfrac{1}{2}(e\kappa_0)^2/h$, with an additional factor $1/2$ to signal the Majorana nature of the quasiparticles.

At finite $E=eV_1$ we must use the energy-dependent transmission matrix \eqref{tNSNfinal}, which gives
\begin{equation}
G_{12}(E)=\tfrac{1}{2}\frac{e^2}{h}N_{\text{Landau}}\left(\cos\theta+\cos\theta'+\cos^2\theta+\cos^2\theta'\right).\label{G12finiteV}
\end{equation}
Substituting Eq.\ \eqref{QNresult} for $\cos\theta$ and $\cos\theta'$ at $k_z$ and $k'_z$, given as a function of $E$ by Eq.\ \eqref{kzEdef}, we find
\begin{equation}
G_{12}(E)=G_{12}(0)\left(1-\frac{\Delta_0^2 E}{(\beta^2-\Delta_0^2)^{3/2}}+{\cal O}(E^2)\right).\label{G12finiteVworkedout}
\end{equation}

The energy dependence of the differential conductance comes entirely from the energy dependence of the effective charge: At $E=0$ the electron-like and hole-like chiral Landau levels have precisely opposite effective charge $\pm Q_{\rm eff}$, but for $E\neq 0$ the effective charges differ in absolute value by an amount $\propto dk_z/dE$.

\subsection{Shot noise}
\label{sec_Pshot}

At temperatures small compared to the applied voltage $V_2$, the time dependent fluctuations in the current $I_2$ are due to shot noise. The formula for the shot noise power is \cite{Ana96}
\begin{equation}
P_{12}=\frac{e^3 V_1}{h}\,{\rm Tr}\,({\cal T}_+-{\cal T}_-^2). \label{P12def}
\end{equation}
This can again be written in terms of the Pauli matrix $\tau_z$ and evaluated using Eq.\ \eqref{tzeroenergy},
\begin{equation}
P_{12}=\frac{e^3 V_1}{h}\left(1-\tfrac{1}{2}\kappa^2-\tfrac{1}{2}\kappa^4\right).\label{P12result}
\end{equation}
The shot noise vanishes when $\kappa\rightarrow 1$, it is fully due to the charge renormalization.

The Fano factor $F$, the dimensionless ratio of shot noise power and average current, results as
\begin{equation}
F=\frac{P_{12}}{eV_1 G_{12}}=\frac{1}{\kappa^2}-\tfrac{1}{2}(1+\kappa^2).\label{Fdef}
\end{equation}

\subsection{Thermo-electricity}

Because of the energy dependence of the effective charge, a temperature difference $\delta T$ between contacts 1 and 2 will produce an electrical current $I_{12}=\alpha_{12}\delta T$ in addition to a heat current. The thermo-electric coefficient $\alpha_{12}$ is given by \cite{Siv86}
\begin{equation}
\alpha_{12}=\frac{\pi^2}{3e}k_{\rm B}^2 T_0\lim_{E\rightarrow 0}\frac{d}{dE}G_{12}(E).\label{alpha12def}
\end{equation}
Substitution of Eq.\ \eqref{G12finiteVworkedout} gives
\begin{align}
\alpha_{12}&=-\frac{\pi^2}{3e}k_{\rm B}^2 T_0 G_{12}(0)\frac{\Delta_0^2}{(\beta^2-\Delta_0^2)^{3/2}}\nonumber\\
&=-g_0 e\kappa_0^2 N_{\rm Landau}\frac{\Delta_0^2}{(\beta^2-\Delta_0^2)^{3/2}}\nonumber\\
&=-g_0 e N_{\rm Landau}\frac{(\Delta_0/\beta)^2}{(\beta^2-\Delta_0^2)^{1/2}}.\label{alpha12result}
\end{align}

\section{Numerical simulations}
\label{sec_numerical}

To test these analytical results, we have carried out numerical calculations in a tight-binding model of the Weyl superconductor with a vortex lattice.

\begin{figure}[tb]
\centerline{\includegraphics[width=0.9\linewidth]{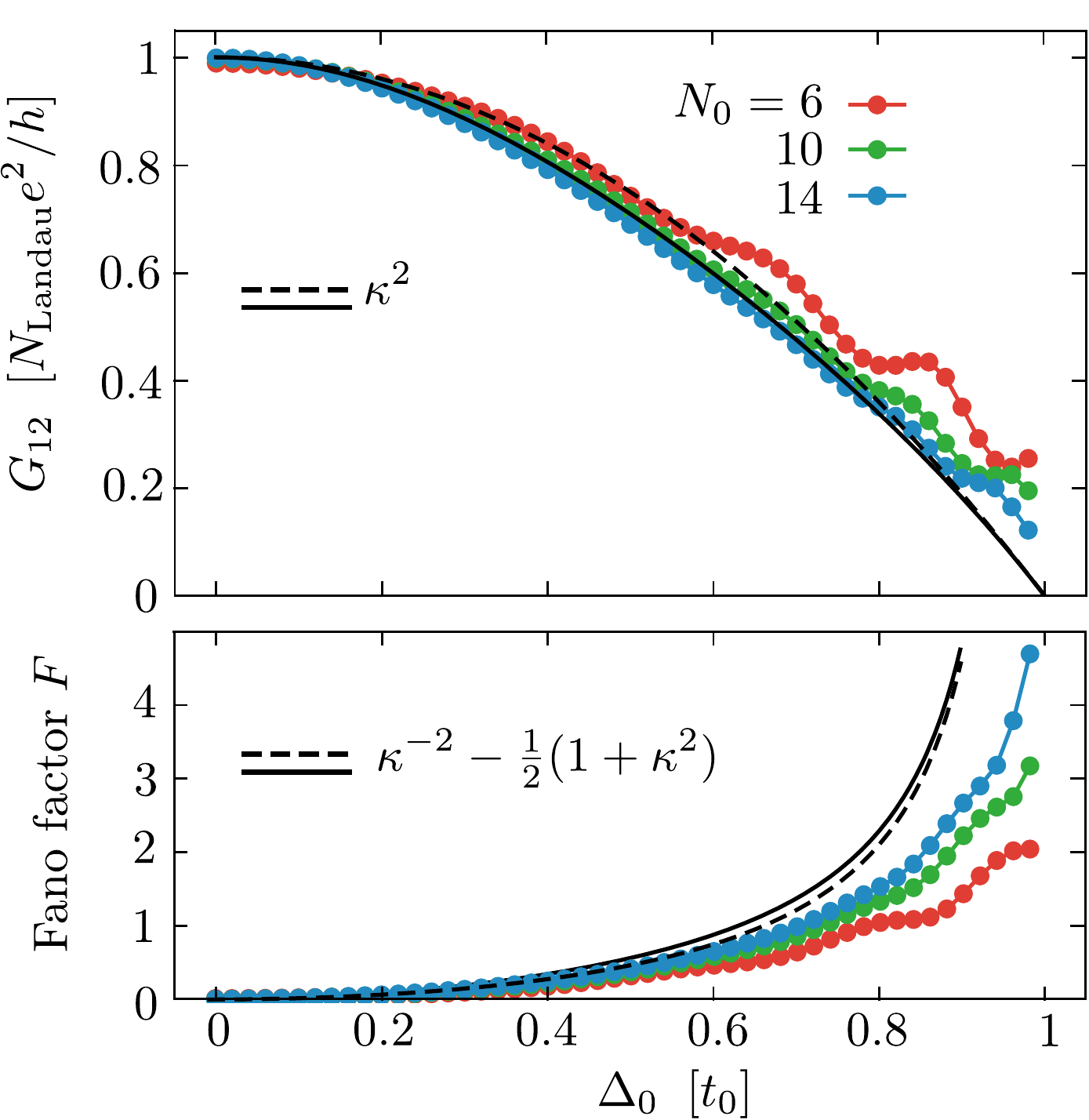}}
\caption{Data points: Electrical conductance (top panel) and Fano factor (bottom panel) in the superconducting vortex lattice (lattice constant $d_0$), as a function of the pair potential $\Delta_0$ at fixed magnetization $\beta=1$, calculated from the tight-binding model (lattice constant $a_0$) for different lattice constant ratios $N_0=d_0/a_0$. The black curves are the analytical predictions from the charge renormalization factor $\kappa$, both in the approximation of a linearized dispersion (black dashed curve, $\kappa=\kappa_0=\sqrt{1-\Delta_0^2/\beta^2}$) and for the full nonlinear dispersion (black solid).
}
\label{fig_G_F_results}
\end{figure}

\begin{figure}[tb]
\centerline{\includegraphics[width=0.9\linewidth]{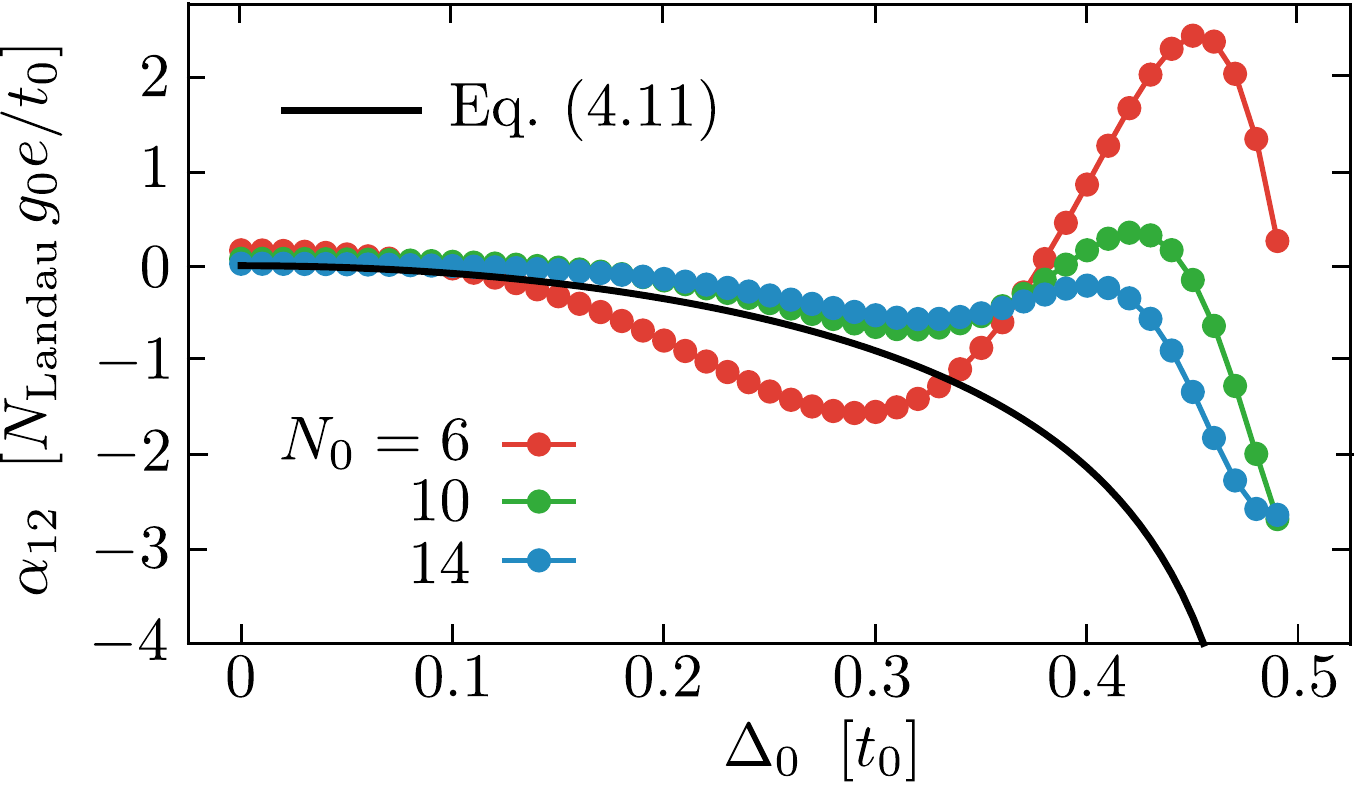}}
\caption{Dependence on $\Delta_0$ for $\beta=0.5$ of the thermo-electric coefficient \eqref{alpha12def}, calculated from the infinite-system analytics (black solid curve) or obtained from finite-size numerics (colored data points). 
}
\label{fig_thermo}
\end{figure}

\begin{figure}[tb]
\centerline{\includegraphics[width=0.9\linewidth]{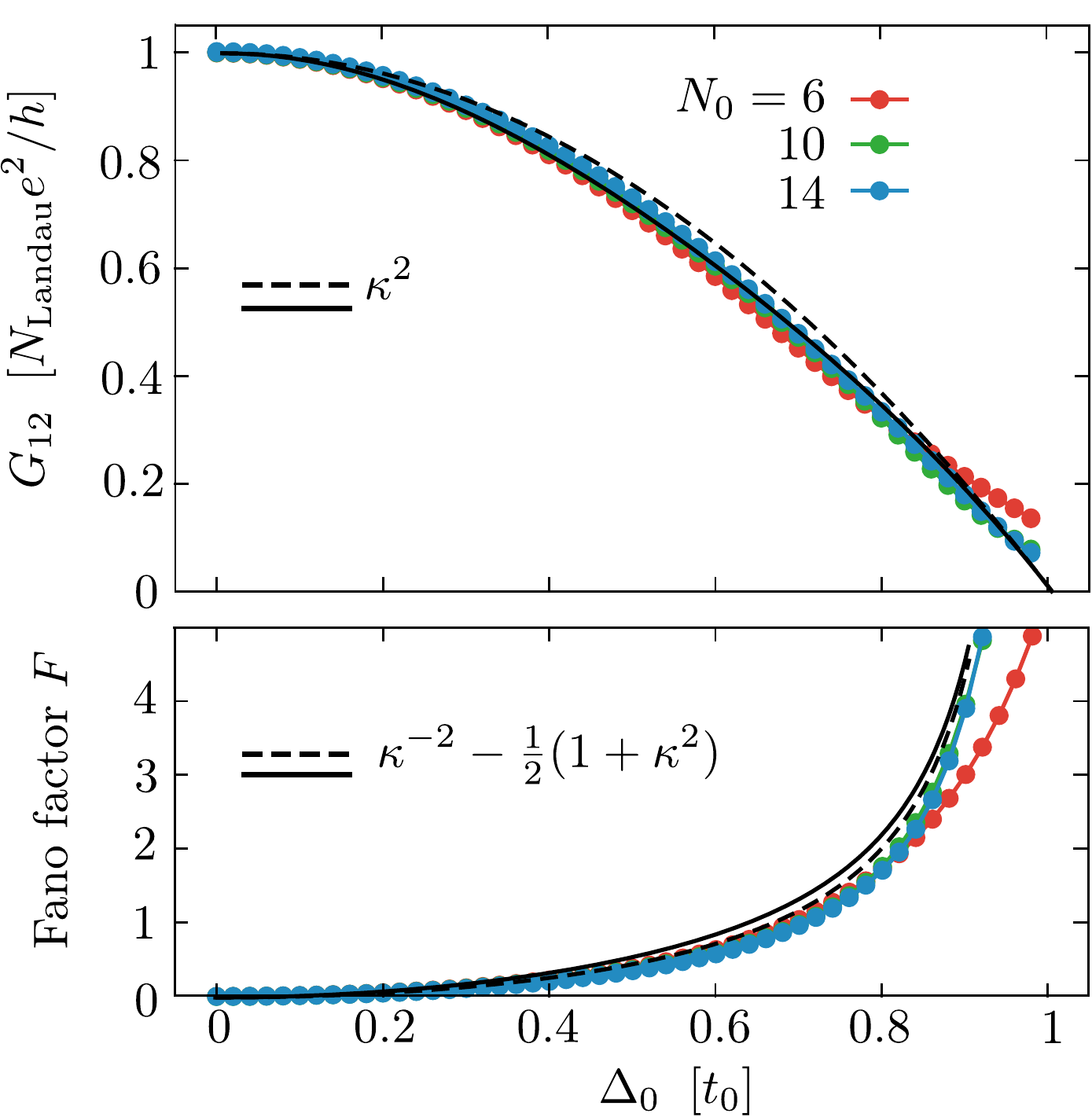}}
\caption{Same as Fig.\ \ref{fig_G_F_results}, but for a magnetization $\beta$ that is perpendicular rather than parallel to the magnetic field $B$.
}
\label{fig_G_F_results_perp}
\end{figure}

\subsection{Tight-binding Hamiltonian}
\label{sec_TBH}

The Bogoliubov-de Gennes Hamiltonian $H_{\rm S}$ in the superconducting region $0<z<L$ is
\begin{subequations}
\label{HBdGSdef}
\begin{align}
H_{\rm S}={}&
\begin{pmatrix}
H_0(\bm{k}+e\bm{A})&\Delta\\
\Delta^\ast&-\sigma_y H_0^\ast(-\bm{k}+e\bm{A})\sigma_y
\end{pmatrix},\label{HBdGSdefa}\\
H_0(\bm{k})={}&t_0{\sum_{\alpha=x,y,z}}\left[\tau_z\sigma_\alpha\sin k_\alpha a_0+\tau_x\sigma_0(1-\cos k_\alpha a_0)\right]\nonumber\\
&+\beta\tau_0\sigma_z-\mu\tau_0\sigma_0.\label{HBdGSdefb}
\end{align}
\end{subequations}
The cubic lattice constant of the tight-binding model is $a_0$ and $t_0$ is the nearest-neigbor hopping energy. In what follows we will set $a_0$ and $t_0$ both equal to unity. 

In the strong-type-II limit the magnetic field $\bm{B}=B_0\hat{z}$ penetrates the superconductor uniformly, with vector potential $\bm{A}=(-B_0y,0,0)$. The absolute value $\Delta_0$ of the pair potential $\Delta=\Delta_0 e^{i\phi}$ can also be taken uniform, assuming that the size $\xi_0=\hbar v_{\rm F}/\Delta_0$ of the vortex core is small compared to the magnetic length $l_m=\sqrt{\hbar/eB_0}$. For the analytical calculations this is the only requirement. For the numerics we also take $\xi_0$ small compared to the tight-binding discretization length $a_0$, and then ensure that a vortex core (where the phase field is undefined) does not coincide with a lattice point. This implies that $a_0$ is large compared to the atomic lattice constant (which itself must be much smaller than $\xi_0$).  

The vortices are arranged on a square lattice in the $x$--$y$ plane, lattice constant $d_0=N_0a_0$, with two $h/2e$ vortices in a unit cell. The number
\begin{equation}
N_0=(a_0^2eB_0/h)^{-1/2}\label{N0def}
\end{equation}
is set at an integer value. The phase $\phi(\bm{r})$ winds around the vortex cores $\bm{R}_n$ according to
\begin{equation}
\nabla\times\nabla{\phi}=2\pi\hat{z}\textstyle{\sum_n}\delta(\bm{r}-\bm{R}_n).\label{curlgradphi}
\end{equation}

In the normal metal leads $z<0$, $z>L$ we have $\Delta_0\equiv 0$ and a large chemical potential $\mu_{\rm N}$, so only modes with a large longitudinal momentum $k_z$ couple to the superconductor. We effectuate the $\mu_{\rm N}\rightarrow\infty$ limit by removing the transverse $x,y$ couplings in the leads, resulting in the Hamiltonian \cite{note1}
\begin{equation}
    H_{\rm N}=\nu_z\tau_{z}\sigma_{z}\sin{k_{z}}+\nu_z\tau_x\sigma_{0}(1-\cos{k_{z}}).\label{HBdGNdef}
\end{equation}
The gauge-invariant discretization of the Hamiltonian \eqref{HBdGSdef} in the magnetic Brillouin zone is detailed in Ref.\ \onlinecite{Pac18}. The scattering matrix is calculated using the Kwant code \cite{kwant}.

\subsection{Results}
\label{sec_results}

Results for the conductance and shot noise are shown in Fig.\ \ref{fig_G_F_results}, as a function of $\Delta_0$ for $\beta=1$, $\mu=0$. The analytical predictions \eqref{G12result} for the conductance and \eqref{Fdef} for the Fano factor are given by the black curves. As a check, for these curves we have also calculated the charge renormalization factor $\kappa$ from the full sinusoidal dispersion, without making the small-$\bm{k}$ expansion of Eq.\ \eqref{Edispersion} --- the difference with $\kappa_0=\sqrt{1-\Delta_0^2/\beta^2}$ is small.

To assess finite-size effects in the numerics we show results for different values of the ratio $N_0=d_0/a_0$ of magnetic unit cell and tight-binding unit cell. As expected, the agreement between numerics and analytics improves with increasing $N_0$, for $\Delta_0/\beta$ not close to unity. (At $\Delta_0=\beta$ the spectrum becomes gapless and the low-energy analytics breaks down.)

These are results at the Fermi level, $E=0$. The energy dependence of the conductance determines the thermo-electric coefficient \eqref{alpha12def}. We show numerical results for $\alpha_{12}\propto dG_{12}/dE$ in Fig.\ \ref{fig_thermo}, for a smaller $\beta=0.5$ to reduce the oscillations that disappear only slowly with increasing $N_0$.

\subsection{Test for isotropy of the charge renormalization}
\label{sec_isotropy}

So far we assumed that the internal magnetization $\beta$ is parallel to the external magnetic field in the $z$-direction. This assumption is needed for our low-energy analytics, but numerically we can take an arbitrary angle between the magnetization $\bm{\beta}=(\beta_x,\beta_y,\beta_z)$ and the magnetic field, by replacing the term $\beta\tau_0\sigma_z$ in the Hamiltonian \eqref{HBdGSdefb} with $\tau_0\,\bm{\beta}\cdot\bm{\sigma}$. Results for $\bm{\beta}=(\beta,0,0)$, so for a magnetization perpendicular to the magnetic field, are shown in Fig.\ \ref{fig_G_F_results_perp}. There is no qualitative difference with Fig.\ \ref{fig_G_F_results} for the parallel configuration, the quantitative difference is that the finite-size effects are smaller.

\section{Conclusion}
\label{sec_conclude}

In summary, we have shown how the charge renormalization $e\mapsto \kappa e$ of Weyl fermions in a superconducting vortex lattice modifies the electrical and thermo-electrical transport properties. 

In the electrical conductance, the current per vortex is reduced by a factor $\tfrac{1}{2}\kappa^2$ --- a prefactor $1/2$ because of the Majorana nature of the quasiparticles and a factor $\kappa^2$ because of the effective charge. At the Weyl point $\kappa\rightarrow\kappa_0=\sqrt{1-\Delta_0^2/\beta^2}$ depends on the ratio of the superconducting gap $\Delta_0$ and the separation $2\beta$ of the Weyl points of opposite chirality. 

The charge-squared renormalization of the electrical conductance is a simple result, but perhaps not what one might have guessed by analogy with the fractional quantum Hall effect, where a $1/3$ fractional charge reduces the conductance by $1/3$ rather than $1/9$. The key difference is that here the quasiparticles are not in an eigenstate of charge; the charge renormalization is due to quantum fluctuations, which give uncorrelated reductions by $\kappa\times \kappa$ at entrance and exit. These quantum fluctuations of the charge are also responsible for the large shot noise power that we have found, with a diverging Fano factor \eqref{Fdef} in the limit $\kappa\rightarrow 0$.

The energy dependence of the charge renormalization implies that charge transport parallel to the magnetic field $B$ goes hand-in-hand with heat transport. As a result, a nonzero thermo-electric coefficient $\alpha_{12}$ along the field lines appears in a chiral Landau level --- something that would not be possible in the normal state: The Landau level contributes an energy-independent number of propagating modes along $B$ (one mode per flux quantum) and the chirality suppresses backscattering, so the energy derivative in Eq.\ \eqref{alpha12def} would vanish in the normal state.  

There is much recent interest in thermo-electricity of Weyl fermions in a Landau level \cite{Ski18,Koz19,Zha19,Han19}, but that refers to currents perpendicular to $B$. Our findings show that charge renormalization in a Weyl superconductor provides a mechanism for a nonzero effect parallel to the field lines.

In our calculations we have assumed a clean system, without impurity scattering. However, we expect the transport properties to be robust against non-magnetic disorder, which in the effective low-energy Hamiltonian \eqref{Heffdef} would enter as a term proportional to $\sigma_z$ that does not couple Landau levels of opposite chirality.

\acknowledgments

This project has received funding from the Netherlands Organization for Scientific Research (NWO/OCW), from the T\"{U}B\.{I}TAK grant No.\ 114F163, and from the European Research Council (ERC) under the European Union's Horizon 2020 research and innovation programme.

\clearpage

\appendix

\begin{widetext}
\section{Calculation of transport properties from the continuum limit of the tight-binding model}

In the tight-binding model of Sec.\ \ref{sec_TBH} the wave matching at the normal-superconductor (NS) interface is implemented by a nearest-neighbor coupling on a square lattice of the Hamiltonians \eqref{HBdGSdef} in S to \eqref{HBdGNdef} in N. Microscopically this results in different matching conditions on the wave function than the matching conditions \eqref{Psimatch} from the analytical treatment of Sec.\ \ref{sec_transmission}. In this Appendix we check that the continuum limit of the tight-binding model still gives the same results for the transport properties as obtained in Sec.\ \ref{sec_transport} from the main text. For simplicity, we set $\mu=0$ and restrict our considerations to $E=0$.

\subsection{Matching condition}

The linearized Hamiltonian for the normal metal reads
\begin{equation}
H_{\rm N} = \nu_z\tau_z\sigma_z k_z
\end{equation}
and for the superconductor it reads
\begin{equation}
H_{\rm S} = \begin{pmatrix} \tau_z\bm\sigma\cdot(\bm{k}-e\bm{A})+\beta\sigma_z & \Delta_0e^{i\phi} \\
\Delta_0 e^{-i\phi} & -\tau_z\bm\sigma\cdot(\bm{k}+e\bm{A})+\beta\sigma_z 
\end{pmatrix}\,.
\end{equation}
The particle current operator is the same for both the normal metal and the superconductor,
\begin{equation}
J_p = \nu_z\tau_z\sigma_z\,,
\end{equation} 
therefore, at the NS interfaces the matching condition
\begin{equation}
\Psi(z=0_-) = \Psi(z=0_+)\,,\qquad 
\Psi(z=L+0_-) = \Psi(z=L+0_+),\label{matchingcond}
\end{equation}
will respect the particle current conservation. This matching condition corresponds to the continuum limit of the tight-binding model of the interface.

As done in Sec. \ref{sec_effcharge}, we start by examining a single NS interface at $z=L$, with a superconductor at $z<L$ and a normal metal at $z>L$. In contrast to the situation described in the main text, the incident modes in the superconductor from Eq. \eqref{PsiKspinor},
\begin{equation}\label{llmaintext}
\begin{split}
&\Psi_{\rm S}\propto \cos(\theta/2)|\mbox{++$-$}\rangle_{\nu\tau\sigma}+\sin(\theta/2)|\mbox{$-$+$-$}\rangle_{\nu\tau\sigma},\\
&\Psi'_{\rm S}\propto \cos(\theta'/2)|\mbox{+++}\rangle_{\nu\tau\sigma}-\sin(\theta'/2)|\mbox{$-$++}\rangle_{\nu\tau\sigma}.
\end{split}
\end{equation}
can no longer be continuously matched to an outgoing state in the normal lead
\begin{equation}
\Psi_{\rm N} \in \operatorname{span}( |\mbox{+++}\rangle_{\nu\tau\sigma},
|\mbox{$-$+$-$}\rangle_{\nu\tau\sigma},
|\mbox{$--$+}\rangle_{\nu\tau\sigma},
|\mbox{+$--$}\rangle_{\nu\tau\sigma}
)\,.
\end{equation}
To satisfy the matching condition \eqref{matchingcond}, an evanescent wave is excited in the superconductor. (There are no evanescent modes in the normal metal.) Because all the incident modes reside in the $\tau=+1$ sector and different $\tau$ sectors are decoupled, in what follows we will focus on $\tau=+1$ sector, and omit the $\tau$ component of the spinor.

\subsection{Evanescent modes}
The evanescent modes are the eigenstates of the effective low energy Hamiltonian \eqref{Heffdefa} with $\operatorname{Im}(k_z)<0$. In this section we will show how to construct them.

We first investigate the spectrum of $H_+$ for $k_z$ in the vicinity of $-K$: $k_z = -K+\delta k_z$, $M(-K+\delta k_z)=\kappa(-K)\delta k_z +\mathcal{O}(\delta k_z^2)$, $\kappa(-K)\equiv\kappa>0$, $\theta'=\theta(-K)$:
\begin{equation}
H_+ = \begin{pmatrix}
\kappa \delta k_z & D \\
D^\dag & -\kappa \delta k_z
\end{pmatrix}\,,\qquad
D = -i\partial_x - \partial_y + e\mathcal{A}_{+,x}-i e\mathcal{A}_{+,y}.
\end{equation}
The states at zero energy satisfy
\begin{equation}\label{zeromodeeq}
H_+ \psi = 0\,.
\end{equation}
Acting with $H_+$ from the left on both sides of the equation yields
\begin{equation}
H_+^2 \psi = 0\,,\quad
H_+^2 = \begin{pmatrix}
(\kappa \delta k_z)^2 + DD^\dag & 0\\
0 & (\kappa \delta k_z)^2 + D^\dag D
\end{pmatrix}\,,
\end{equation}
therefore the two components of the state $\psi=(\psi_1, \psi_2)^T$ must be be eigenstates of $DD^\dag$ and $D^\dag D$ respectively, with the same eigenvalue. Suppose we can find the eigenstates of $D^\dag D$:
\begin{align}
D^\dag D \phi_n &= \epsilon_n \phi_n\,,\quad \epsilon_n>0\,,\quad n=1,2\dots
\end{align}
Note that $\epsilon_n = 0$ is not allowed as shown in Ref.\ \onlinecite{Pac18}, and $\epsilon_n\geq 0$ because they are the eigenvalues of a square of a Hermitian operator. The operator $DD^\dag$ has one zero-mode: $\psi_0$, $D^\dag\psi_0 = 0$. The remainder of the spectrum of $DD^\dag$ can be obtained by acting with operator $D$ on wavefunctions $\phi_n$,
\begin{equation}
D D^\dag (D \phi_n) = \epsilon_n D\phi_n\,.
\end{equation}
This means that the sought wavefunction can be written as
\begin{equation}
\psi = \begin{pmatrix}
\alpha D\phi_n \\ \beta \phi_n
\end{pmatrix}\quad \text{or}\quad 
\psi = \begin{pmatrix}
\psi_0 \\ 0
\end{pmatrix}\,.
\end{equation}
The second possibility corresponds to the propagating zeroth Landau level. Therefore, we will now focus on the first possibility. Substituting the wavefunction of this form into eigenvalue equation \eqref{zeromodeeq} yields
\begin{equation}
\begin{pmatrix}
\kappa\delta k_z & D \\
D^\dag & -\kappa\delta k_z
\end{pmatrix}\begin{pmatrix}
\alpha D\phi_n \\ \beta \phi_n
\end{pmatrix} = 0\,,
\end{equation}
which gives us
\begin{equation}
\kappa\delta k_z = \pm i\sqrt{\epsilon_n} \,,\qquad \beta = \mp i\alpha\sqrt{\epsilon_n}.
\end{equation}
We choose the lower sign in the solution in order to satisfy the condition $\operatorname{Im}(k_z)<0$. With this we can obtain the evanescent modes of the full Hamiltonian $\mathcal{H}$,
\begin{subequations}\label{evanescentmodes}
	\begin{equation}
	\Psi =
	e^{(-i K+\sqrt{\epsilon_n})z}
	e^{\frac{1}{2}i\theta'\nu_y\sigma_z} \lvert +\rangle_\nu \begin{pmatrix}
	D\phi_n \\ i\sqrt{\epsilon_n} \phi_n
	\end{pmatrix}_\sigma
	=
	\begin{pmatrix}
	D\phi_n \cos\tfrac{\theta'}{2} \\
	i\sqrt{\epsilon_n} \phi_n \cos\tfrac{\theta'}{2} \\
	-D\phi_n \sin\tfrac{\theta'}{2} \\
	i\sqrt{\epsilon_n} \phi_n \sin\tfrac{\theta'}{2}
	\end{pmatrix} e^{(-i K+\sqrt{\epsilon_n}) z}\,,
	\end{equation}
	where the spinor on the right hand side is written in the basis $\lvert ++\rangle_{\nu\sigma}, \lvert +-\rangle_{\nu\sigma}, \lvert -+\rangle_{\nu\sigma}, \lvert--\rangle_{\nu\sigma}$.
	The evanescent modes corresponding to $k_z$ around $+K$ can be obtained by acting with charge conjugation operator on $\Psi$,
	\begin{equation}
	\Psi' = \nu_y\sigma_y\mathcal{K} \Psi =
	\begin{pmatrix}
	i\sqrt{\epsilon_n} \phi_n^* \sin\tfrac{\theta'}{2} \\
	-D^*\phi_n^* \sin\tfrac{\theta'}{2} \\
	-i\sqrt{\epsilon_n} \phi_n^* \cos\tfrac{\theta'}{2} \\
	-D^* \phi_n^* \cos\tfrac{\theta'}{2} \\
	\end{pmatrix} e^{(i K+\sqrt{\epsilon_n})z}\,.
	\end{equation}
\end{subequations}
We can also obtain the two chiral Landau levels
\begin{subequations}
	\begin{equation}\label{llappendixa}
	\Psi_S' =
	e^{-i K z}
	e^{\frac{1}{2}i\theta'\nu_y\sigma_z}
	\lvert +\rangle_\nu 
	\begin{pmatrix}
	\psi_0 \\ 0
	\end{pmatrix} = \begin{pmatrix}
	\psi_0\cos\tfrac{\theta'}{2} \\
	0 \\
	-\psi_0 \sin\tfrac{\theta'}{2} \\
	0
	\end{pmatrix}e^{-i K z},
	\end{equation}
	\begin{equation}
	\Psi_S =
	\nu_y\sigma_y\mathcal{K}\Psi_S =
	\begin{pmatrix}
	0 \\
	\psi_0 \sin\tfrac{\theta'}{2} \\
	0 \\
	\psi_0\cos\tfrac{\theta'}{2} 
	\end{pmatrix}e^{i K z},
	\end{equation}
\end{subequations}
which correspond to states in Eq. \eqref{llmaintext}. From now on we will drop the prime at $\theta'$ and define $\theta=\theta(-K)$.

\subsection{Transmitted wave}

We will now consider an incoming wave which is $\Psi_{\rm S}'$ -- the chiral Landau level at zero energy with momentum $k_z = -K$, cf. Eq. \eqref{llappendixa}. The solution for $\Psi_{\rm S}$ can be obtained using particle-hole symmetry. We want to find a superposition of evanescent modes Eq. \eqref{evanescentmodes} such that its profile at the interface $z=L$,
\begin{equation}
\Psi_{\rm eva} = \sum_n 
\alpha_n\begin{pmatrix}
D\phi_n \cos\tfrac{\theta}{2} \\
i\sqrt{\epsilon_n} \phi_n \cos\tfrac{\theta}{2} \\
-D\phi_n \sin\tfrac{\theta}{2} \\
i\sqrt{\epsilon_n} \phi_n \sin\tfrac{\theta}{2}
\end{pmatrix} +
\alpha_n'\begin{pmatrix}
i\sqrt{\epsilon_n} \phi_n^* \sin\tfrac{\theta}{2} \\
-D^*\phi_n^* \sin\tfrac{\theta}{2} \\
-i\sqrt{\epsilon_n} \phi_n^* \cos\tfrac{\theta}{2} \\
-D^* \phi_n^* \cos\tfrac{\theta}{2} \\
\end{pmatrix}\,,
\end{equation}
will satisfy the boundary condition
\begin{equation}
\Psi_{\rm S}' + \Psi_{\rm eva} = \Psi_{\rm N}'.
\end{equation}
Writing it down explicitly we get
\begin{equation}\label{boundarycondexpl}
\begin{pmatrix}
\psi_0 \cos\tfrac{\theta}{2} \\
0 \\
- \psi_0\sin\tfrac{\theta}{2} \\
0
\end{pmatrix} + \sum_n 
\alpha_n\begin{pmatrix}
D\phi_n \cos\tfrac{\theta}{2} \\
i\sqrt{\epsilon_n} \phi_n \cos\tfrac{\theta}{2} \\
-D\phi_n \sin\tfrac{\theta}{2} \\
i\sqrt{\epsilon_n} \phi_n \sin\tfrac{\theta}{2}
\end{pmatrix} +
\alpha_n'\begin{pmatrix}
i\sqrt{\epsilon_n} \phi_n^* \sin\tfrac{\theta}{2} \\
-D^*\phi_n^* \sin\tfrac{\theta}{2} \\
-i\sqrt{\epsilon_n} \phi_n^* \cos\tfrac{\theta}{2} \\
-D^* \phi_n^* \cos\tfrac{\theta}{2} \\
\end{pmatrix} = \begin{pmatrix}
g_1 \\ 0 \\ 0 \\ g_2
\end{pmatrix}\,,
\end{equation}
where $g_1$, $g_2$ are some functions of $\bm r = (x,y)$. If we project both sided of the equation on the second and third component of the spinor, we obtain
\begin{equation}
\begin{split}
0 &= \sum_n \alpha_n	i\sqrt{\epsilon_n} \phi_n \cos\tfrac{\theta}{2}
- \alpha'_n D^*\phi_n^* \sin\tfrac{\theta}{2} ,\\
\psi_0 \sin\tfrac{\theta}{2} &=
\sum_n -\alpha_n D\phi_n \sin\tfrac{\theta}{2} 
- \alpha'_n i\sqrt{\epsilon_n} \phi_n^* \cos\tfrac{\theta}{2}\,,
\end{split}
\end{equation}
or equivalently
\begin{equation}
\begin{split}
\sum_n \alpha_n i\sqrt{\epsilon_n} \phi_n
&= \sum_n\alpha'_n D^*\phi_n^* \sin\tfrac{\theta}{2}/\cos\tfrac{\theta}{2},\\
\sum_n\alpha_n D\phi_n &=
\sum_n
- \alpha'_n i\sqrt{\epsilon_n} \phi_n^* \cos\tfrac{\theta}{2}/ \sin\tfrac{\theta}{2}
-\psi_0\,.
\end{split}
\end{equation}
Substituting this back into Eq. \eqref{boundarycondexpl} and projecting it on the first and fourth component we get
\begin{align}
\begin{pmatrix}
g_1 \\ g_2
\end{pmatrix} &= \sum_n 
\alpha_n\begin{pmatrix}
D\phi_n \cos\tfrac{\theta}{2} \\
i\sqrt{\epsilon_n} \phi_n \sin\tfrac{\theta}{2}
\end{pmatrix} +
\alpha_n'\begin{pmatrix}
i\sqrt{\epsilon_n} \phi_n^* \sin\tfrac{\theta}{2} \\
-D^* \phi_n^* \cos\tfrac{\theta}{2}
\end{pmatrix} + \begin{pmatrix}
\psi_0\cos\tfrac{\theta}{2} \\ 0
\end{pmatrix} \nonumber\\ &=
\sum_n 
\alpha_n'\begin{pmatrix}
- i\sqrt{\epsilon_n} \phi_n^* \cos^2\tfrac{\theta}{2}/ \sin\tfrac{\theta}{2} \\
D^*\phi_n^* \sin^2\tfrac{\theta}{2}/\cos\tfrac{\theta}{2}
\end{pmatrix} +
\alpha_n'\begin{pmatrix}
i\sqrt{\epsilon_n} \phi_n^* \sin\tfrac{\theta}{2} \\
-D^* \phi_n^* \cos\tfrac{\theta}{2}
\end{pmatrix} + \begin{pmatrix}
-\psi_0\cos\tfrac{\theta}{2} \\ 0
\end{pmatrix} + \begin{pmatrix}
\psi_0\cos\tfrac{\theta}{2}\\0\end{pmatrix} \nonumber\\
&=
\sum_n 
\alpha_n'\begin{pmatrix}
i\sqrt{\epsilon_n} \phi_n^* (\sin\tfrac{\theta}{2} - \cos^2\tfrac{\theta}{2}/ \sin\tfrac{\theta}{2})\\
-D^* \phi_n^*(\cos\tfrac{\theta}{2} - \sin^2\tfrac{\theta}{2}/\cos\tfrac{\theta}{2})
\end{pmatrix} \nonumber\\
&=
\sum_n 
\alpha_n'\begin{pmatrix}
-i\sqrt{\epsilon_n} \phi_n^* \kappa/\sin\tfrac{\theta}{2}\\
-D^* \phi_n^* \kappa/\cos\tfrac{\theta}{2}
\end{pmatrix}\,,
\end{align}
therefore, the transmitted wave has the form
\begin{equation}
\Psi_N' = \sum_n 
\alpha_n'\begin{pmatrix}
-i\sqrt{\epsilon_n} \phi_n^* \kappa/\sin\tfrac{\theta}{2}\\ 0\\0\\
-D^* \phi_n^* \kappa/\cos\tfrac{\theta}{2}
\end{pmatrix}.
\label{evaresult}
\end{equation}

\subsection{Charge transfer}
Since there is no reflection, we normalize the outgoing wave such that the outgoing particle current is 1:
\begin{align}
\langle \Psi_{\rm N} \rvert J_p \lvert \Psi_{\rm N}\rangle &= \sum_{nm}
\alpha_n'^*
\begin{pmatrix}
-i\sqrt{\epsilon_n} \phi_n^* \kappa/\sin\tfrac{\theta}{2}\\ 0\\0\\
-D^* \phi_n^* \kappa/\cos\tfrac{\theta}{2}
\end{pmatrix}^\dag
\alpha_m'\begin{pmatrix}
-i\sqrt{\epsilon_m} \phi_m^* \kappa/\sin\tfrac{\theta}{2}\\0\\0\\
-D^* \phi_m^* \kappa/\cos\tfrac{\theta}{2}
\end{pmatrix} \nonumber\\
&=
\sum_{nm}
\alpha_n'^*	\alpha_m'[
\sqrt{\epsilon_n} \sqrt{\epsilon_m} (\phi_n^*)^\dag \phi_m^* \kappa/\sin\tfrac{\theta}{2}\kappa/\sin\tfrac{\theta}{2} + (\phi_n^*)^\dag (D^*)^\dag D^* \phi_m^* \kappa/\cos\tfrac{\theta}{2} \kappa/\cos\tfrac{\theta}{2}
]\nonumber\\
&=
\sum_{n}
|\alpha_n'|^2 \epsilon_n [
\kappa^2/\sin^2\tfrac{\theta}{2} + \kappa^2/\cos^2\tfrac{\theta}{2}
] \nonumber\\
&= (\kappa^2/\sin^2\tfrac{\theta}{2} + \kappa^2/\cos^2\tfrac{\theta}{2}) \sum_{n}
|\alpha_n'|^2 \epsilon_n  \overset{!}{=} 1\,. \label{evanormalization}
\end{align}

Albeit coefficients $\alpha_n'$ cannot be determined in a closed form, the information we obtained in Eqs. \eqref{evaresult} and \eqref{evanormalization} is sufficient to calculate the transport properties. In particular, the transmitted electric charge is given by
\begin{align}
\langle \Psi_{\rm N}' \rvert e\nu_z \lvert \Psi_{\rm N}'\rangle &= 
e\sum_{n}
|\alpha_n'|^2 \epsilon_n [
\kappa^2/\sin^2\tfrac{\theta}{2} - \kappa^2/\cos^2\tfrac{\theta}{2}
] \nonumber\\
&= 
e\frac{\kappa^2/\sin^2\tfrac{\theta}{2} - \kappa^2/\cos^2\tfrac{\theta}{2}}{\kappa^2/\sin^2\tfrac{\theta}{2} + \kappa^2/\cos^2\tfrac{\theta}{2}} \nonumber\\
&= e\frac{1/\sin^2\tfrac{\theta}{2} - 1/\cos^2\tfrac{\theta}{2}}{1/\sin^2\tfrac{\theta}{2} + 1/\cos^2\tfrac{\theta}{2}} = \cos^2\theta/2 - \sin^2\theta/2 = e\kappa = Q_{\rm eff}\,,
\end{align}
which is the same result as $Q_{\rm N}'$ obtained in the main text in Eq. \eqref{Qeffdef}. The transmitted wave for the incident mode $\Psi_{\rm S}$ is $\Psi_{\rm N} = \nu_y\sigma_y\mathcal{K}\Psi_{\rm N}'$ (as required by the particle hole symmetry). Therefore, the corresponding transmitted charge is $\langle \Psi_{\rm N}\rvert e\nu_z \lvert \Psi_{\rm N}\rangle=-\kappa e = -Q_{\rm eff}$.

\subsection{Transport properties}

An analogous analysis can be performed for the interface at $z=0$, yielding the corresponding incident waves in the $z<0$ metallic lead: $\Phi_{\rm N}$ and $\Phi_{\rm N}'$, which couple perfectly to the chiral Landau levels $\Psi_{\rm S}$, $\Psi_{\rm S}'$ respectively. This yields a transmission matrix
\begin{equation}
t = e^{i K L} \lvert\Psi_{\rm N}\rangle\langle\Phi_{\rm N}\rvert + e^{-i K L}
\lvert\Psi_{\rm N}'\rangle\langle\Phi_{\rm N}'\rvert\,,
\end{equation}
like in Eq. \eqref{tNSNE}, with the difference that the modes in the superconductor can no longer be written explicitly in a closed form.
Still we can compute the thermal conductance
\begin{align}
G_{\rm thermal} &= \frac{1}{2}g_0 N_{\rm Landau} \operatorname{Tr}{ t^\dag t} = g_0 \frac{e\Phi}{h}\,,
\label{Gthermappendix}
\end{align}
where we used that $\langle \Psi_N|\Psi_N'\rangle = 0$, as required by the unitarity of the scattering matrix. We can also compute the electric conductance
\begin{align}
G_{12} &= \frac{e^2}{h}N_{\rm Landau} \operatorname{Tr}{\frac{\nu_z+1}{2} t\nu_{z}t^{\dagger}} \nonumber \\ &= \frac{e^2}{h}N_{\rm Landau}\left(
\langle\Psi_{\rm N}\rvert\nu_z\lvert\Psi_{\rm N}\rangle
\langle\Phi_{\rm N}\rvert\frac{\nu_z+1}{2}\lvert\Phi_{\rm N}\rangle
+
\langle\Psi_{\rm N}'\rvert\nu_z\lvert\Psi_{\rm N}'\rangle
\langle\Phi_{\rm N}'\rvert\frac{\nu_z+1}{2}\lvert\Phi_{\rm N}'\rangle\right) \nonumber \\ &= \frac{e^2}{h}N_{\rm Landau}\frac{1}{2}\Big{[}\kappa(1+\kappa)-\kappa(1-\kappa)\Big{]} = \frac{(e\kappa)^2}{h}\frac{e\Phi}{h}\,, \label{G12appendix}
\end{align}
Where we used the fact that 
\begin{equation}
\langle \Psi_{\rm N}\rvert \nu_z\lvert\Psi_{\rm N}'\rangle =
\int d\bm{r} \begin{pmatrix}
g_2^*(\bm r) \\ 0 \\ 0 \\ g_1^*(\bm r)
\end{pmatrix}^\dag \nu_z
\begin{pmatrix}
g_1(\bm r) \\ 0 \\ 0 \\ g_2(\bm r)
\end{pmatrix} = \int d\bm{r} [g_2(\bm r)g_1(\bm r) -g_1(\bm r)g_2(\bm r)] = 0\,,
\end{equation}
and similarly $\langle \Phi_{\rm N} \rvert \nu_z\lvert \Phi_{\rm N}'\rangle = 0$

The thermal and electric conductance obtained in Eqs. \eqref{Gthermappendix} and \eqref{G12appendix} are identical to the results obtained in the main text: Eqs. \eqref{Gthermalresult} and \eqref{G12result}. Furthermore, a similar calculation shows that the shot noise power is also given by the same formula as in the main text: Eq. \eqref{P12result}. This confirms that the tight-binding model is equivalent in the continuum limit to the analytics.

\end{widetext}

\end{document}